# Pioneer Women in Chaos Theory

Frank Y. Wang
LaGuardia Community College of the City University of New York

The general public has been made aware of the research field of *Chaos* by the book of that title by James Gleick.[1] Since the publication of that best seller in 1987, the term "chaos" has become a trendy word, and the title of the leading chapter "butterfly effect" is a household name. While the idea of chaos seemed to emerge recently, it actually arose from the prize-winning work of one of the greatest mathematicians of the late nineteenth century—Henri Poincaré (1854–1912). Poincaré's 1890 memoir on the three-body problem was the result of his entry in King Oscar II of Sweden's 60th birthday competition. The Russian mathematician, Sonya Kovalevskaya (1850–1891), then a professor of the University of Stockholm, was consulted in the offering of the prize. About the same time, she finished her own celebrated work on the motion of a rigid body.[2] From today's point of view, with the benefit of 20/20 hindsight, the works of Poincaré and Kovalevskaya hinted at the failure of Newtonian determinism by using Newton's own laws. The implication is that chaos is ubiquitous.

The butterfly effect[3] is often ascribed to the meteorologist Edward Lorenz (1917–2008), and the numerical simulations of three simple equations named after him are favorite images for textbooks and magazines. Freeman Dyson pointed out that Mary Lucy Cartwright (1900–1998) and John Edensor Littlewood (1885–1977) studied the chaos-like behavior in the equation for a radio amplifier known as the van der Pol oscillator during World War II, predating Lorenz's work by decades.[4] In Lorenz's own book *The Essence of Chaos*,[5] he elaborated the crucial influence of the study by Littlewood and Cartwright. Why did the work of Lorenz precipitate an enthusiastic public response while the work of Cartwright and Littlewood 20 years earlier did not? Dyson believed that the change of style in mathematics is one of the reasons. The new mathematics is visual rather than analytical: chaos theory flourished and became popular because computers were able to simulate motions accurately and display them dramatically. This explanation might also apply to the situation of Kovalevskaya, whose analytical approach to the rigid body problem was a *tour de force*, but is under appreciated by contemporary mathematicians and physicists. The purpose of this paper is to revisit Kovalevskaya's and Cartwright's mathematical works and place them in proper historical perspective in the development of chaos theory. To encourage the new visual style of mathematical thinking, the author has developed supplemental materials using commonly available computer software such as *Maple*, *Mathematica*, and *Matlab*, so that students can learn chaos theory and examine the pioneer works from a modern perspective.[6]

---

[1] James Gleick, *Chaos: Making a New Science*, Viking, New York, 1987.
[2] S. V. Kovalevskaya, "Sur le problem de la rotation d'un corps solide autour d'um point fixe," *Acta Mathematica*, **XII**, 177–232 (1889).
[3] Technically, the "butterfly effect" might be loosely translated as "sensitive dependence on initial conditions." For some history of this phrase, see Robert C. Hilborn, "Sea gulls, butterflies, and grasshoppers: A brief historyof the butterfly effectin nonlinear dynamics," *American Journal of Physics*, **72**, 425–427 (2004).
[4] Freeman Dyson, "Book Reviews," *American Mathematical Monthly*, **103**, 610–612 (1996); Dyson, "Mary Lucy Cartwright, Chaos Theory" in *Out of the Shadows: Contributions of Twentieth-Century Women to Physics*, edited by Nina Byers and Gary Williams, Cambridge University Press, New York, 2006.
[5] Edward N. Lorenz, *The Essence of Chaos*, University of Washington Press, Seattle, 1993.

Periodic motion has a strong hold in the history of science. Galileo noticed that the motion of a chandelier hanging in a cathedral oscillated with a constant period, and Kepler analyzed voluminous astronomical data and revealed regular elliptical orbits for the planets. Newton was able to show that the regular swinging of a pendulum and the trajectory of planets follow from his laws of motion. His second law ($F = ma$) is in the form of differential equations, which relate not just quantities, but also the rates at which those quantities changes. Since Newton, differential equations have been proven to be the most effective ways to express theories of physics; they constitute a powerful tool to predict natural phenomena surrounding us, like the vibrations of a string, the ripples on the surface of a pond, and the forever evolving patterns of the weather. The revolution in scientific thought that culminated in Newton led to a vision of the universe as some gigantic mechanism, functioning like clockwork.

The great success of Newton's theory had an unfortunate consequence in education. By solving textbook problems of pendulum oscillations and planetary motion using routine methods, students got the impression that mechanics is about writing the differential equations of motion and finding the analytical solutions. They were led to an erroneous impression that all motions were periodic based on idealized simplifications, and that more difficult problems involved merely technical refinements, which can be solved using more powerful computers.

In spite of the great elegance and simplicity of Newton's treatment of the motion of celestial bodies (with subsequent improvements by Bernoulli and others), the investigation cannot be regarded as complete. The solar system consists of planets, their moons, comets, and asteroids, but the canonical treatment is to consider two bodies only and ignore the rest. For a long term prediction of planetary motion, it is necessary to consider additional bodies in the system. The three-body problem was a natural first step and captured the imagination of mathematicians since Newton. Yet other than some restricted situations, no general solution has been obtained. It was considered to be so important that in 1885 King Oscar II of Sweden offered a prize of 2500 crowns[7] for the solution to the three-body problem. Like others before him, Poincaré failed to solve the equations. But unlike others he solved the problem in a very different sense: he proved that the equations could not be solved. Through attacking the three-body problem, Poincaré had laid the foundations for the modern approach to dynamical systems using topology.

Another important application of Newtonian mechanics occurs in the study of the motion of a rigid body. The period formula ($T = 2\pi\sqrt{l/g}$) and the solution (a sinusoidal function) that students learn early in school is based on the assumption that the bob is a point oscillating on a plane. A rigid body problem is a generalization from a point to an extended body rotating in a three-dimensional space. Like the three-body problem, the rigid body problem also attracted the attention of many great mathematicians. Euler studied the case that gravity is indifferent, and Lagrange focused on a symmetric rigid body. There was no progress for 100 years after Lagrange's work (1788), until Kovalevskaya made a breakthrough. Her result is best described in a letter that she sent to Götta Mittag-Leffler in 1886:[8]

[6] Contact the author at fwang@lagcc.cuny.edu for supplemental materials; some electronic documents are available at http://faculty.lagcc.cuny.edu/fwang.
[7] For comparison, the annual salary of a top professor such as Gösta Mittag-Leffler was 7000 crowns.

> Dear Sir,
>
> I thank you for your invitation for tomorrow, and I shall come with pleasure. It is a question of integrating the following differential equations,
>
> $$A\frac{dp}{dt} = (B-C)qr + z_0\gamma' - y_0\gamma'' \qquad \frac{d\gamma}{dt} = q\gamma'' - r\gamma'$$
> $$B\frac{dq}{dt} = (C-A)rp + x_0\gamma'' - z_0\gamma \qquad \frac{d\gamma'}{dt} = r\gamma - p\gamma''$$
> $$C\frac{dr}{dt} = (A-B)pq + y_0\gamma - x_0\gamma' \qquad \frac{d\gamma''}{dt} = p\gamma' - q\gamma$$
>
> Up to now they have been integrated only in 2 cases: (1) $x_0 = y_0 = z_0 = 0$ (the case of Poisson and Jacobi); (2) $A = B,\ x_0 = 0$ (the Lagrange case).
>
> I have found the integral also in the case $A = B = 2C,\ z_0 = 0$, and I can show that these 3 cases are the only ones in which the general integral [i.e. the solution for every set of initial values of the variables] is a single-valued analytic function of time having no singularities but poles for finite values of $t$. …

The differential equations in this letter are the Euler equations, which are explained in supplemental material. Kovalevskaya's letter highlights her two achievements. First she set a new case of the motion of a rigid body, and gave a solution in terms of hyperelliptic functions.[9] Furthermore, she proved that with the exception of three cases it is impossible to find a general solution of the problem of motion of a rigid body in terms of analytic functions.

The achievements of Kovalevskaya and Poincaré were their realization that in general one cannot find analytical solutions that would describe the position of the rigid bodies or planets at all times. Traditionally, a differential equation is solved by finding a function that satisfied the equation; a trajectory is then determined by starting the solution with a particular initial condition. Before the discoveries of Poincaré and Kovalevskaya, it was thought that a nonlinear system would always have a solution; we just needed to be clever enough to find it. But such a view is wrong. Kovalevskaya and Poincaré showed that no matter how clever we are, we will not be able to solve most of the differential equations. The belief in determinism, that the present state of the world determines the future precisely, was shattered.

[8] The letter was written in French; this English translation is by Roger Cooke in *The Mathematics of Sonya Kovalevskaya*, Springer-Verlag, New York, 1984.

[9] An integral of the form $\int_0^x du/\sqrt{P(u)}$ is called an elliptic integral if $P(u)$ is a polynomial of either the third or the forth degree. If $P(u)$ is of degree higher than the forth, the integrall is called Abelian or hyperelliptic.

Although Kovalevskaya's portrait can be found in many mathematics departments' galleries, most people have the vaguest idea about her contributions. One possible reason is Dyson's "analytical versus visual" argument. Among the three analytically solvable rigid body cases, the Kovalevskaya case is the most complicated and the most difficult one. She followed the analytical tradition of Lagrange, which is intrinsically tedious. In Lagrange's *Mécanique Analytique*, he attempted to mold the whole system of mechanics into analytical mathematical equations, and was proud of the fact that one would not find a single diagram in his work. Kovalevskaya's calculations extended over 50 pages, involving elliptic functions unfamiliar to today's scholars and students. Furthermore, Kovalevskaya's mathematical accomplishments and her relationship to the mathematical community of her time have been subjected to distortion, particularly by E. T. Bell's classic *Men of Mathematics*, in which he implied that Kovalevskaya's mentor Karl Weierstrass did most of the work. Without a question, Weierstrass (1815–1897) was the most influential mathematician in Kovalevskaya's career. He is regarded as the father of modern analysis; his definitions of limit and derivative are what today's students learn in their calculus course. It is true that Weierstrass suggested Kovalevskaya work on the problem, but Kovalevskaya was the first person to apply the ideas of the theory of functions of a complex variable developed by Cauchy, Riemann, and Weierstrass to the solution of a dynamical problem. The passages in Bell's book seem haughtily patronizing and freighted with inappropriate innuendo.[10]

Even in the mathematical physics circle, while the Euler[11] and Lagrange cases are included in commonly used classical mechanics textbooks such as those by Goldstein,[12] by Landau and Liftshitz,[13] and by Arnold,[14] Kovalevskaya's theorem and her treatment of her own case are unknown to students and even professors. To prepare students for future studies, educators must emphasize the limited solvability in mechanical systems. Instead of finding the individual solutions of differential equations, the modern emphasis has been shifted from finding a local solution to a global solution by applying geometrical and topological techniques. It is understandable that Kovalevskaya's intricate transformation of the integrals of motion to elliptic functions might be of interest to aficionados only, but students can apply numerical methods to visualize the actual motion of the rigid body problem. In Figure 1, we show some numerical results of the Kovalevskaya case. This two-way approach of using abstract topological methods on the one hand and insightful computer experiments on the other is central to the methodology employed when studying nonlinear dynamical systems.

[10] For example, on page 426 of *Men of Mathematics* Bell wrote: "Man's ingratitude to man is a familiar enough theme; Sonja now demonstrated what a woman can do in that line when she puts her mind to it. She did not answer her old friend's letter for two years although she knew had had been unhappy and in poor health. The answer when it did come was rather a letdown. …"

[11] In Kovalevskaya's letter above, she referred to the Euler case as the "case of Poisson and Jacobi."

[12] H. Goldstein, C. P. Poole and J. L. Safko, *Classical Mechanics*, 3rd ed., Addison Wesley, San Francisco, 2002.

[13] L. D. Landau and E. M. Lifshitz, *Mechanics*, 3rd ed., Pergamon Press, Oxford, 1976.

[14] V. I. Arnold, *Mathematical Methods of Classical Mechanics*, 2nd ed., Springer-Verlag, New York, 1989.

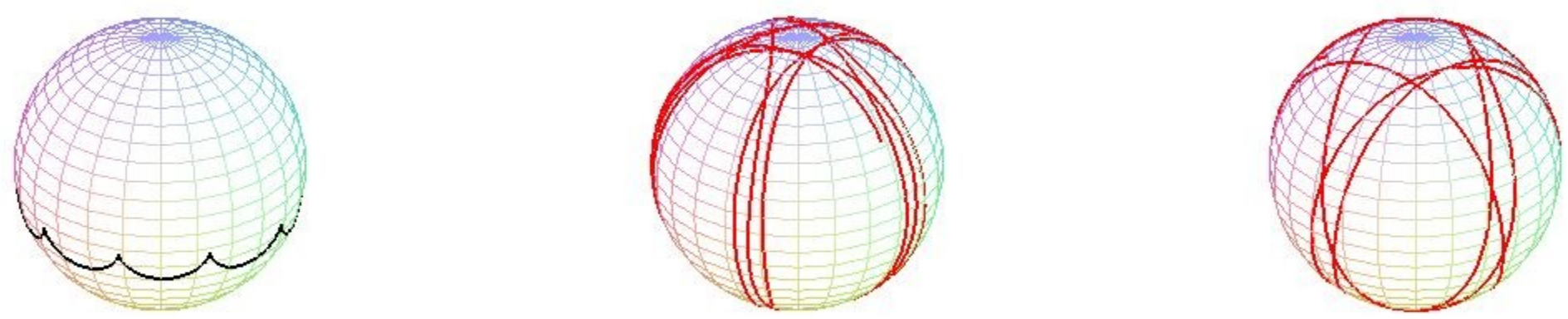

**Figure 1:** The motion of a rigid body of the Kovalevskaya case by tracing a curve of the intersection of the body axes with the surface of a sphere of unit radius about the fixed point.

After Poincaré's death, mathematicians soon lost track of the important ideas surrounding his discovery. The major exceptions were A. Lyapunov in Russia and G. D. Birkhoff in America, who continued to advance classical mechanics in the early twentieth century. In England a gulf had formed between the pure and applied mathematics; this situation was reflected in G. H. Hardy's essay *A Mathematician's Apology*, in which he declared that pure mathematics must be useless. In 1938, the Second World War was pending, and Britain was belatedly preparing to defend itself. The most important technical instrument for the defense of Britain was radar. But at that time, the deployment of radar was greatly hampered by the lack of reliable high-powered radio amplifiers. The Radio Research Board of the U. K. Department of Scientific and Industrial Research issued a call for help, requesting the "really expert guidance" of pure mathematicians with "certain types of non-linear differential equations involved in the technique of radio engineering." Mary Cartwright responded to the appeal. Prior to it, she studied under Hardy in the late 1920s, and obtained her D.Phil. from Oxford in 1930. One of her examiners was Littlewood. Littlewood had a rather eccentric personality, and it was notoriously difficult to work with him. Mary Cartwright intuitively recognized the topological features of the van der Pol oscillator, and was able to bring it to Littlewood's attention.[15]

Cartwright and Littlewood began studying the van der Pol equation just before the War, and the collaboration lasted approximately a decade. They published four joint papers; their paper entitled "On Non-Linear Differential Equations of the Second Order: I. The Equation $\ddot{y} - k(1-y^2)\dot{y} + y = b\lambda k\cos(\lambda t + a)$, $k$ Large" in *Journal of the London*

[15] S. L. McMurran and J. J. Tattersall, "The Mathematical Collaboration of M. L. Cartwright and J. E. Littlewood," *American Mathematical Monthly*, **103**, 833–845 (1996).

*Mathematical Society* is the most frequently quoted one.[16] It might be difficult to read the original paper, but with the aid of commonly used software students can understand their statement through visualization. Cartwright and Littlewood noticed that for some parameter ranges, two distinct stable subharmonic motions were obtained, and for some other parameter ranges the Poincaré section possessed a great variety of structures. They were puzzled by their result, but were confident about its validity because van der Pol had found a similar phenomenon experimentally.[17] Specifically, van der Pol reported that "often an irregular noise is heard in the telephone receivers before the frequency jumps to the next lower value" in a letter to *Nature*.[18] In the caption of Figure 2 the reader can find technical explanations, and it is easy to employ *Mathematica* to simulate the sound that was heard by van der Pol. The observation of Cartwright and Little implied the existence of a "strange attractor," which is neither a point nor a curve in the Poincaré section.

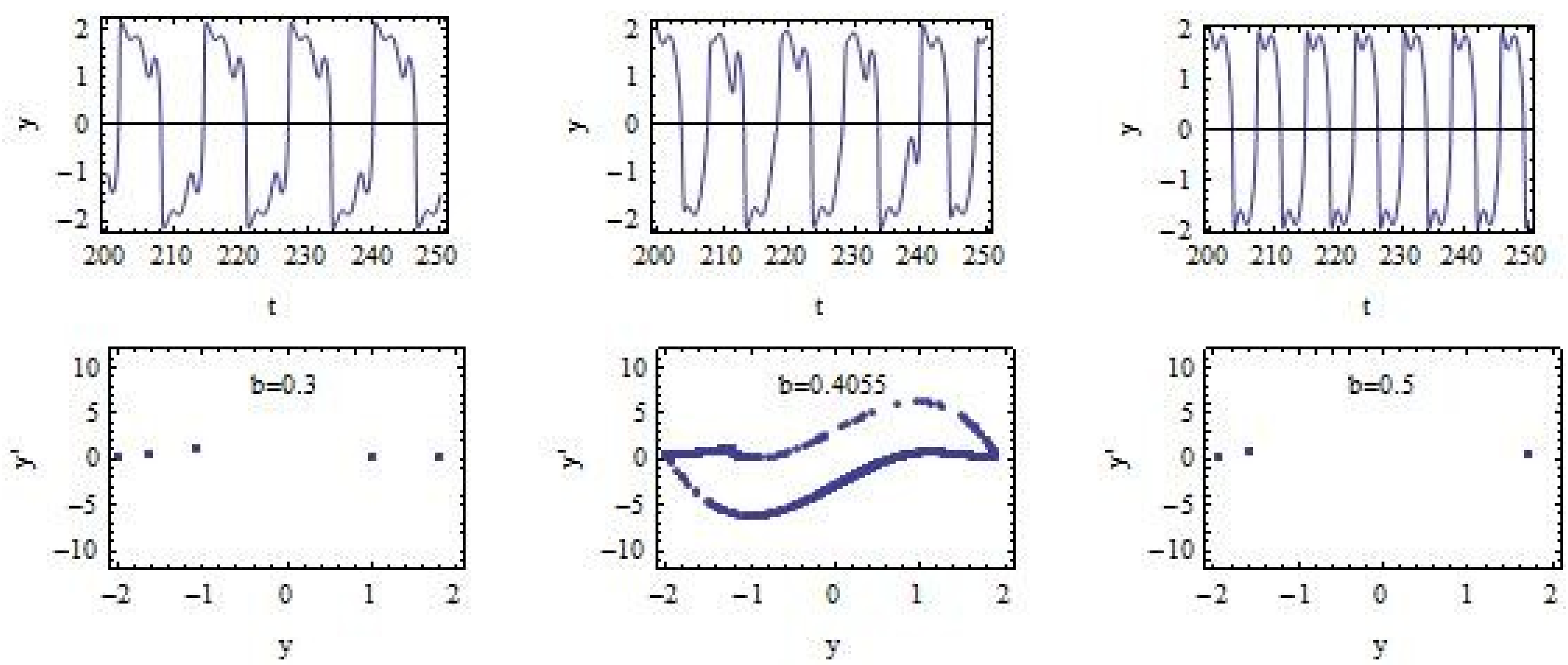

**Figure 2**: Numerical solutions to $\ddot{y} - k(1-y^2)\dot{y} + y = b\lambda k\cos(\lambda t + a)$, with $k = 5$, $\lambda = 2.466$, $a = 0$, and $b$ values indicated in the figure. The top row shows the time series, or the plot *y* versus *t*. The bottom shows the Poincaré section, which is the value of *y* and its slope *y'* each time when $t$ is an integer multiple of $\varpi = 2\pi/\lambda$. A periodic motion of least period $n\varpi$ is called a subharmonic of order $n$. For $b = 0.3$ and $b = 0.5$, there are a finite number of periodic motion (5 and 3, respectively). For $b = 0.4055$, the Poincaré section shows an infinite number of periodic motion of a great variety of "structures" described by Cartwright and Littlewood; such a structure is called a strange attractor in modern terms.

Although the paper of Cartwright and Littlewood announced groundbreaking results, the year 1945 was a bad time to publish. Paper in England was scarce and few copies of the Journal containing Cartwright's paper were printed; the paper attracted little attention. In 1949, Cartwright was invited to lecture on nonlinear differential equations at Princeton

[16] M. L. Cartwright and J. E. Littlewood, "On Non-Linear Differential Equations of the Second Order: I. The Equation $\ddot{y} - k(1-y^2)\dot{y} + y = b\lambda k\cos(\lambda t + a)$, $k$ Large," *Journal of the London Mathematical Society*, **20**, 180–189 (1945).

[17] In a footnote, Cartwright and Littlewood wrote "our faith in our results was at one time sustained only by the experimental evidence that stable sub-harmonics of two distinct orders did occur."

[18] B. van der Pol and J. van der Mark, "Frequency Demultiplication," *Nature*, **120**, 363–364 (1927).

University, but her results attracted little attention either. Cartwright later learned that if her host, Solomon Lefschetz, stopped asking questions for five minutes, he was asleep. Fortunately, the significance of Cartwright's work was understood by Norman Levinson, who gave a simpler analysis.[19] Their works led to Stephen Smale's introduction of the horseshoe map.

The reputation of Steve Smale (born in 1930) was not confined to mathematics. He said that his best work had been done "on the beaches of Rio."[20] He used this beaches reference to mock Washington politicians. In 1960, Smale was in Rio de Janeiro as a postdoctoral fellow sponsored by the U. S. National Science Foundation. His antiwar activities in the 1960s created a furor in Washington. In 1966, while the U. S. House Un-American Activities Committee was trying to subpoena him, he was heading for Moscow to attend the International congress of Mathematics. There he received the Fields Medal, the most prestigious prize in mathematics. Subsequently, questions were raised about his having used U. S. taxpayers' money for this research on the beaches of Rio. Smale recalled that shortly after his arrival in Rio, Levinson wrote him a letter pointing out a mistake in a conjecture Smale just published. The conjecture was "chaos doesn't exist!" Smale eventually convinced himself that Levinson was correct, as chaos was already implicit in the analyses of Cartwright and Littlewood. Through geometrizing the analytical results of Cartwright, Littlewood and Levinson, Smale invented the horseshoe map (on the beach), the driving mechanism and cornerstone of the whole modern theory of chaos. The idea also led Smale to a solution of Poincaré's Conjecture in dimensions greater than 4, and it was for that work he received the Fields Medal.

A fascinating aspect about chaos is that such a phenomenon arises in diverse fields. From a statistical-mechanical view, heat is the manifestation of randomly jiggling molecules, thus chaos is actually desirable. In early 1950s, Enrico Fermi, John Pasta and Stan Ulam set to study how a crystal evolves toward thermal equilibrium using one of the very first computers, the MANIAC. They numerically integrated the differential equation, and expected a stochastic motion. To their surprise, they found the motion to be highly regular. The Fermi-Pasta-Ulam (FPU) problem, first written up in a Los Alamos Report in May 1955,[21] marked the beginning of both a new field, nonlinear physics (which is of central importance in the theory of chaos), and the age of computer simulations of scientific problems. Fermi died of cancer in 1954, and the report never turned into a peer-reviewed paper. Readers of that report might be puzzled by the authorship credit, which reads "Work done by E. Fermi, J. Pasta, S. Ulam, M. Tsingou, Report written by E. Fermi, J. Pasta, S. Ulam." On page 3, a footnote reads "We thank Miss Mary Tsingou for efficient coding of the problems and for running the computations on the Los Alamos MANIAC machine." Here we encountered another unsung heroine, whose identity remained little-known until a recent article published in *Physics Today*.[22]

---

[19] N. Levinson, "A second-order differential equation with singular solutions," *Annals of Mathematics*, **50**, 127–153 (1949).
[20] S. Smale, "Finding a Horseshoe on the Beaches of Rio," *Mathematical Intelligencer*, **20**, 39–44 (1998).
[21] E. Fermi, J. Pasta, S. Ulam and M. Tsingou, "Studies of Nonlinear Problems," Los Alamos preprint LA-1940.
[22] T. Dauxois, "Fermi, Pasta, Ulam, and a Mysterious Lady," *Physics Today*, **61** (1), 55–57 (2008).

Born in 1928 to a Greek family living in Milwaukee, Wisconsin, Mary Tsingou applied for a position at Los Alamos National Laboratory in 1952 following a suggestion by her advanced differential equations professor (a woman). Fermi, Pasta and Ulam were extremely fortunate to find Mary Tsingou, who programmed the dynamics, ensured its accuracy, and provide graphs of the results.[23] After Fermi's death, Jim Tuck and Mary Menzel (née Tsingou) repeated the original FPU results and provided strong indication that the nonlinear FPU problem might be integrable.[24]

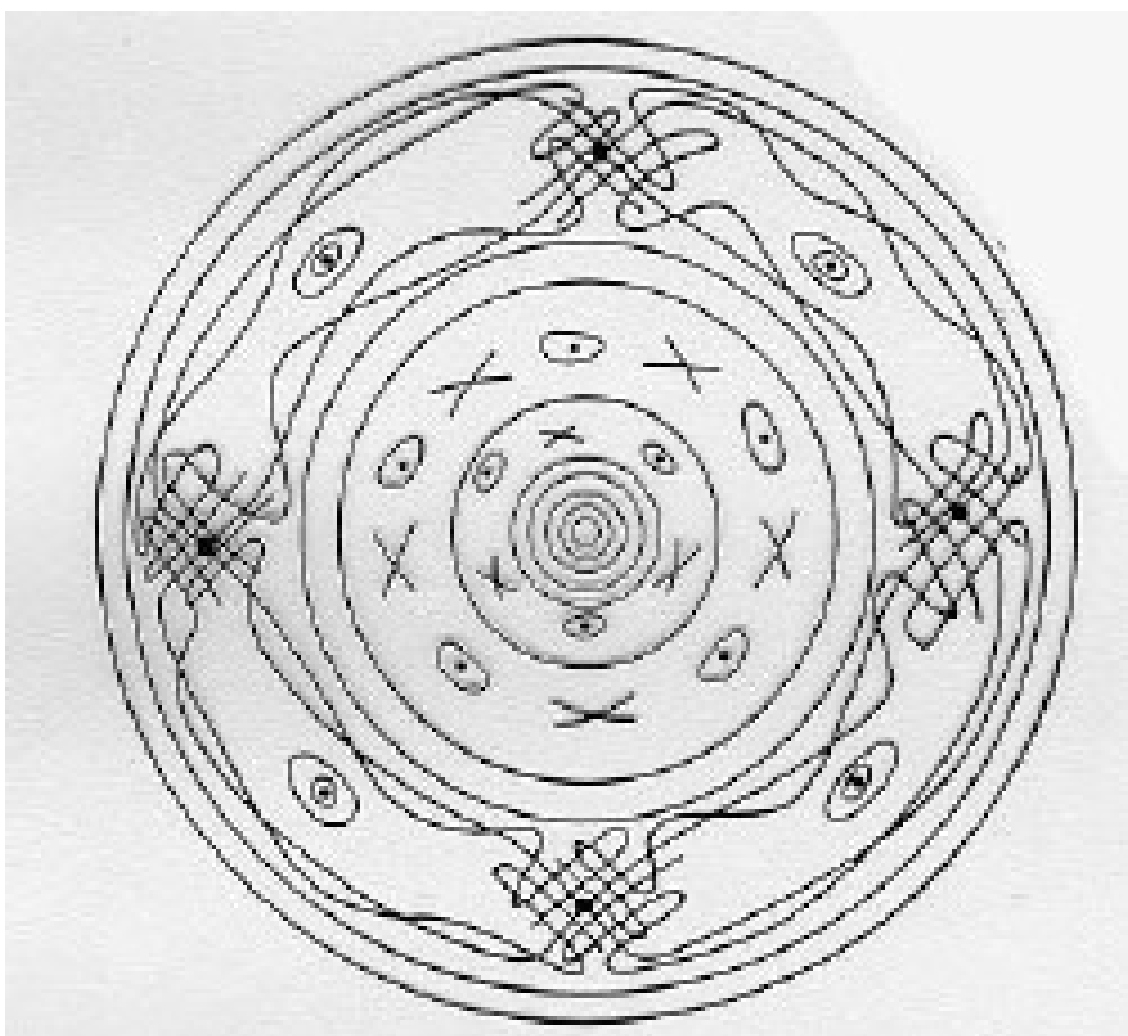

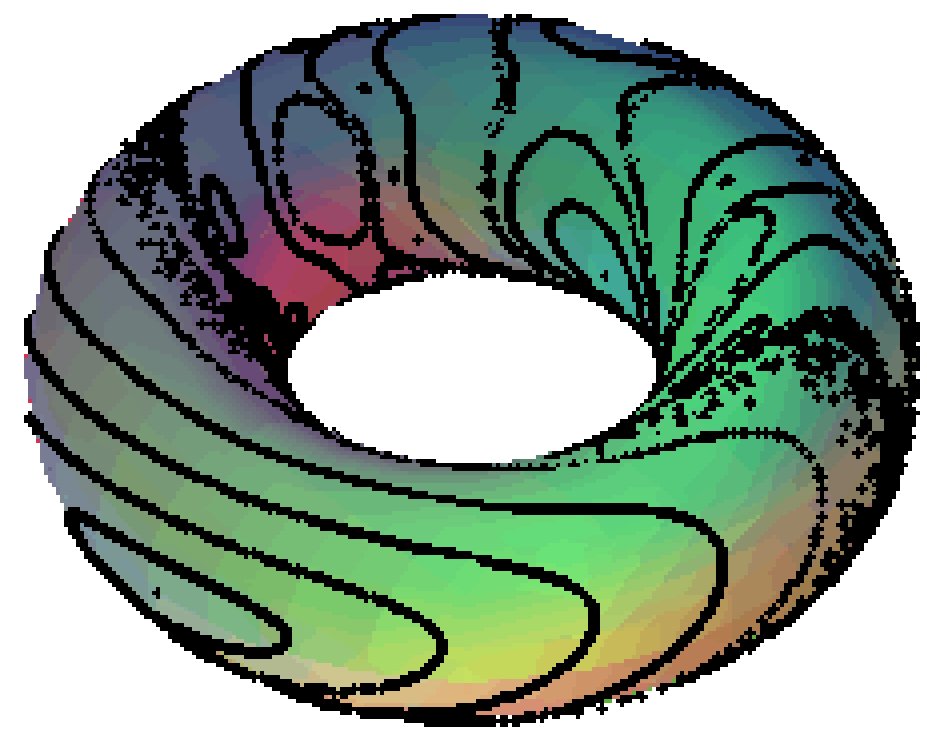

**Figure 3**: Nested KAM tori sketch in *Ergodic Problems of Classical Mechanics* by V. I. Arnold and A. Avez. Some tori persist, and some replaced by chaos. The remaining tori set up barriers which prevent the chaotic trajectories form drifting indiscriminately all over.

The FPU paradox forced physicists to face some of their deepest insecurities.[25] Classical mechanicians thought that the few-body problem was analytically solvable, but Poincaré recognized that the three-body problem could be chaotic. Taking the opposite tack, statistical mechanicians asserted that the many-body problem was stochastic, but the FPU computer calculations revealed the motion to be highly ordered. How to resolve these contradictions? At the International Mathematical Congress of 1954, A. N. Kolmogorov (1903–1987) gave a closing lecture in the Amsterdam Concertgebouw. He suggested a theorem (without proof), which, though no one noticed at the time, could have explained the lack of chaos observed by FPU. The details of Kolmogorov's theorem were later worked out by V. I. Arnold (born in 1937) and Jürgen Moser (1928–1999), now known as the Kolmogorov-Arnold-Moser (KAM) theorem. In modern terms, an integrable system is recognized by invariant tori[26] in phase space. The KAM theorem states that under perturbation, most tori do not vanish, but are only slightly deformed. But some tori do get destroyed and replaced by chaos. The chaos-creating mechanism is similar to the horseshoe map. Indeed, after showing the impossibility of solving the differential equations of planetary motion, Poincaré discovered the inevitability of homoclinic tangles, which turned out to be the trademark of chaos. In his own words,

> The complexity of this figure will be striking, and I shall not even try to draw it. Nothing is more suitable for providing us with an idea of the complex nature of the three-body problem, and all the problems of dynamics in general, where there is no uniform integral and where the Bohlin series are divergent.

[23] Students can repeat Tsingou's task with 15 lines of *Matlab* code; see T. Dauxois, M. Peyrard, and S. Ruffo, "The Fermi-Pasta-Ulam Numerical Experiment: History and Pedagogical Perspectives," *European Journal of Physics*, **26**, S3–S11 (2005).

[24] J. L. Tuck and M. T. Menzel, "The Superperiod of the Nonlinear Weighted String (FPU) Problem," *Advances in Mathematics*, **9**, 399–407 (1972).

[25] J. Ford, "The Fermi-Pasta-Ulam Problem: Paradox Turns Discovery," *Physics Reports*, **213**, 271–310 (1992).

The figure Poincaré would not draw is shown in Figure 3, along with a numerical approximation based on the standard map.[27] One opinion popularly held in the 1950s was that the addition of perturbation would render the system chaotic, but we can observe from Figure 3 that nonlinearity is not enough to guarantee complete chaos, as the remaining tori set up barriers which prevent the chaotic regions from drifting indiscriminately all over phase space. Most orbits are confined to specific regions of phase and remain so indefinitely. One application of the KAM theorem was to address the issue of the stability of the solar system, which has been debated since the eighteenth century. Because the KAM tori set up barriers to prevent chaos from spreading, one might be inclined to call the solar system stable, at least philosophically.

The KAM theorem represents a unified formulation of the complicated structure of regular and irregular trajectories; we hope that this paper will make readers aware of pioneer works done by remarkable women: Kovalevskaya's contribution to the theorem of non-integrability, and Cartwright's insight into the chaos-like topological structure. Pedagogically speaking, a good share of physics and mathematics is still writing differential equations on a blackboard and showing students how to solve them. Although it is impractical to change the conventional curriculum drastically, it is possible to introduce some innovations. The unforced van der Pol equation is a standard topic in differential equations textbooks, and it is a feasible project for students go one step further to investigate the equation Cartwright studied.[28] Integrable systems are the basis for the KAM formulation, and among those few integrable systems, the Kovalevskaya case is arguably the most beautiful one. Kovalevskaya's method can provide a test for integrability,[29] and the Kharlamov's bifurcation analysis[30] of the integral manifolds serves as an excellent example for analyzing the nontrivial nesting of invariant tori. We urge more educators to make Kovalevskaya's legacy accessible to undergraduate students.

---

[26] A most popular example of a two-dimensional torus is a donut-shaped surface.
[27] See "Standard Map on a Torus," http://faculty.lagcc.cuny.edu/fwang/smap/standardmap.html.
[28] This paper might be useful for instructor to design projects for students: U. Parlitz and W. Lauterborn, "Period-Doubling Cascades and Devil's Staircases of the Driven van der Pol Oscillator," *Physical Review*, **36**, 1428–1434 (1987).
[29] M. Tabor, "Modern dynamics and classical analysis," *Nature*, **310**, 277–282 (1984).
[30] M. P. Kharlamov, "Bifurcation of Common Levels of First Integrals of the Kovalevskaya Problem," *Prikl. Matem. Mekhan.* (Journal of Applied Mathematics and Mechanics), **47** (6), 922–930 (1983).